\documentclass[12pt]{iopart}
\usepackage{iopams}
\usepackage{amssymb}
\usepackage[english]{babel}
\usepackage{graphicx}
\usepackage{dcolumn}
\usepackage{bm}
\usepackage{verbatim}
\usepackage{mathrsfs}
\usepackage{cancel}
\usepackage{epstopdf}
\usepackage{color}
\usepackage[usenames,dvipsnames]{pstricks}
\usepackage{epsfig}
\usepackage{pst-grad} 
\usepackage{pst-plot} 
\usepackage{hyperref}
\usepackage{cite}

\newcommand{\ii}{\mathrm{i}}

\newcommand{\be}{\begin{equation}}
\newcommand{\ee}{\end{equation}}
\newcommand{\bea}{\begin{eqnarray}}
\newcommand{\eea}{\end{eqnarray}}

\def\slashchar#1{\setbox0=\hbox{$#1$} 
\dimen0=\wd0 
\setbox1=\hbox{/} \dimen1=\wd1 
\ifdim\dimen0>\dimen1 
\rlap{\hbox to \dimen0{\hfil/\hfil}} 
#1 
\else 
\rlap{\hbox to \dimen1{\hfil$#1$\hfil}} 
/ 
\fi}

\begin{document}

\title{On the Unruh effect, trajectories and information}

\author{Aida Ahmadzadegan$^{1,2}$ and Achim Kempf$^{1,2,3,4}$}
\address{$^1$Department of Applied Mathematics, University of Waterloo, Waterloo, Ontario, N2L 3G1, Canada}
\address{$^2$Perimeter Institute for Theoretical Physics, Waterloo, Ontario N2L 2Y5, Canada}
\address{$^3$Department of Physics and Astronomy, University of Waterloo, Waterloo, Ontario N2L 3G1, Canada}
\address{$^4$Institute for Quantum Computing, University of Waterloo, Waterloo, Ontario, N2L 3G1, Canada}

\ead{aida.ahmadzadegan@uwaterloo.ca}

\begin{abstract}
We calculate the trajectories which maximize the Unruh effect, mode by mode, when given a fixed energy budget for acceleration. We find that Unruh processes are most likely to occur, and therefore potentially best observable, for certain trajectories whose acceleration is not uniform. 
In practice, the precise form of optimal trajectories depends on experimental bounds on how fast the acceleration can be changed. We also show that the Unruh spectra of arbitrarily accelerated observers contain the complete information to reconstruct the observers' trajectories.     



\end{abstract}

\maketitle


\section{Introduction} 

The Unruh effect \cite{Unruh1976} is the phenomenon that accelerated observers experience radiation even when inertial observers experience the vacuum state. In particular, uniformly accelerated observers are predicted to experience thermal radiation that is proportional to the acceleration. The Unruh effect is important in part due to its relation, via the equivalence principle, to the Hawking effect of black holes. Indeed, similar questions can be asked regarding both the Unruh and Hawking effects.

In particular, there is the question to what extent Hawking radiation is modulated as the size of a black hole changes due to infalling matter, and how much information, therefore, Hawking radiation carries about the infalling matter. Here, we will not address this question, see, e.g., \cite{Carney2017,strominger2017,marolf2017,Pati2007,Giddings2006}, but we will study the similar question to what extent Unruh radiation is modulated by changes in the acceleration of a trajectory and how much information, therefore, Unruh radiation carries about the trajectory. Indeed, Unruh radiation generally cannot be expected to be thermal when the detector's acceleration is non-uniform (and/or only temporarily switched on) as this introduces new scales that can affect the spectrum. In fact, it is the nontriviality of the spectrum in the case of non-uniform acceleration that allows one to reconstruct the trajectories, as we will here show.

To this end, we use that, mathematically, the Unruh effect can be traced to the 
fact that wave forms that consists of positive frequencies which vary in time must also possess negative frequencies in its Fourier transform, and vice versa, i.e., that, e.g., a chirp of the form $\e^{\rmi\alpha t^2}$ for $\alpha>0$ contains in its Fourier decomposition finite contributions from plane waves $\e^{\rmi\omega t}$ with $\omega <0$. 
We calculate the trajectories that maximally modulate the Unruh spectrum, and that may, therefore, be useful to improve the experimental accessibility of the Unruh effect. We also show that, for any arbitrarily accelerated trajectory, the excitations and field quanta created by the Unruh effect carry sufficient information to reconstruct the trajectory. 

We begin by considering why there is reason to expect that the Unruh effect can be modulated significantly by variations in the acceleration of the trajectory. 
To this end, we recall that any quantum system that can act as a detector of field quanta must contain a charge in order to couple to the field. As the detector is accelerated so is its charge and the detector will, therefore, generally radiate, i.e, it will excite the quantum field. Vice versa, through the same interaction Hamiltonian, the quantum field can and generally will excite the detector. This can be interpreted as the detector registering Unruh radiation, see e.g., \cite{Birrell1984}.  
For comparison, in the case of accelerated classical charges, effects such as the Abraham Lorentz force \cite{abraham02} or Feynman and Wheeler's radiation resistance \cite{radresistance} are known to be sensitive to variations in acceleration. It is plausible, therefore, that also the Unruh effect can be significantly modulated by variations in a particle detector's acceleration, i.e., by higher-than-second derivatives in the detector's trajectory. For prior work on the Unruh effect for non-uniform acceleration, see, in particular, \cite{Barbado2012a,Lynch2015,Smerlack2015,AasenPRL,Aida2014,Doukas2013,Lin2017,Ostapchuk2011}. 

Our aim here will be to determine to what extent elementary Unruh processes (rather than the overall Unruh effect) can be enhanced or suppressed by suitable choices of non-uniformly accelerated trajectories. By an elementary Unruh process we mean a process in which an accelerated detector creates a field quantum of momentum $\bm{k}$, while transitioning between two energy eigenstates, $\vert E_0\rangle$ and $\vert E_1\rangle$.
We will compare the probability of Unruh processes for these non-uniformly accelerated trajectories to those for the uniformly accelerated trajectory with the same initial and final velocities, i.e., with the same energy budget available for acceleration. 
The reason why we will not trace over the field to obtain the overall Unruh effect, i.e., the reason why we will not integrate over $\bm{k}$, is that these field quanta are observable and carry key information about the trajectory, as we will show.

\section{Non-uniformly accelerated Unruh DeWitt detectors} 
We consider a free scalar field and we model particle detectors as localized first-quantized two-level systems, so-called Unruh-DeWitt (UDW) detectors, with the usual interaction Hamiltonian \cite{DeWitt,Louko2008,Scully1997,Wavepackets,Birrell1984,Alvaro,Pozas2016}:
\bea
\hat{H}_{\textrm{I}}= \lambda\chi(\tau) \hat{\mu}(\tau) \hat{\phi}\left(\textbf{x}(\tau),t(\tau)\right).
\eea
Here, $\tau$ is the proper time of the UDW detector, $\left(t(\tau),\bm{x}(\tau)\right)$ is its trajectory, 
and $\hat{\phi}\left(x(\tau)\right)$ is the field along the trajectory. $\lambda$ is a small coupling constant and $\chi(\tau) \ge 0$ is a window function to switch the detector.
$\hat{\mu}(\tau)$ is the detector's monopole moment operator which is given by $\sigma_z$ as:
\bea \label{monopole}
\hat{\mu}(\tau)=(\hat\sigma^{+} \rme^{\ii\tau \Delta E }+\hat \sigma^{-}\e^{-\ii\tau\Delta E }).
\eea
Here, $\{\sigma^{+}, \sigma^{-}\}$ are ladder operators for a two-sate system such as a spin $1/2$ and $\Delta E = E_1-E_0$ is the proper energy gap between the ground state, $\left| E_0 \right\rangle$, and the excited state, $\left| E_1 \right\rangle$, of the detector. In Minkowski space, the first-order probability amplitude for an UDW detector to register a particle, i.e., to transition from its ground state to its excited state, $\vert E_0\rangle \rightarrow\vert E_1\rangle$ while creating a particle of momentum $\bm{k}\in \mathbb{R}^3$ from the vacuum is given by \cite{Birrell1984}, 

\begin{equation}
\psi_k(\Delta E)= \eta
\int_{-\infty}^{\infty}{\e^{\ii \Delta E~\!\tau}\e^{\ii(\omega_k t(\tau)-\bm{k}.\bm{x}(\tau))}\chi(\tau)} \rmd\tau,
\label{basic}
\end{equation}
where $\eta =\frac{\ii \lambda \left\langle E \right| \hat{\mu}_0 \left| E_0 \right \rangle}{(16 \pi^3 \omega_k)^{1/2}}$.
It will be important for our considerations below that, apart from a constant prefactor, $\psi_k(\Delta E)$ is the Fourier transform of the $\tau$-dependent function $\e^{\ii(\omega_k t(\tau)-\bm{k}.\bm{x}(\tau))}$. Let us define an effective frequency $\bar{\omega}(\tau)$ so that: 

\begin{equation}
    \rme^{\ii\int_0^\tau \bar{\omega}(\tau')\rmd\tau'}=\e^{\ii(\omega_k t(\tau)-\bm{k}.\bm{x}(\tau))}.
\end{equation}
For example, if an always-on detector, $\chi(\tau)\equiv 1$, is on an inertial trajectory through the origin with a velocity $\bm{v}$, then
$\e^{\ii(\omega_k t(\tau)-\bm{k}.\bm{x}(\tau))}
=\e^{\ii \bar{\omega}\tau }$, yielding
\bea
\psi_k(\Delta E) = 2\pi \eta \delta(\Delta E + \bar{\omega}),~~~~ \textrm{with} ~~~~\bar{\omega}= (\omega_k-\bm{k}.\bm{v})(1-\bm{v}^2)^{-1/2}.
\eea
Since $\Delta E\ge 0$ and $\bar{\omega} > 0$, we have $\psi_k(\Delta E)=0$, i.e., an always-on inertial detector will, of course, not become excited. In contrast, $\psi_k(\Delta E)$ is non-vanishing if $\Delta E$ is negative, namely if $\Delta E=-\bar{\omega}$, which is the case of an initially excited detector that decays while emitting a field quantum of momentum $\bm{k}$.   

\section{Concomitant frequencies} 
The amplitude, $\psi_k(\Delta E)$, for an Unruh process to occur, i.e., for an accelerated detector that starts in the ground state (i.e., $\Delta E>0$) to get excited in the Minkowski vacuum can be nonzero only if the Fourier transform of the trajectory-dependent function $\e^{\ii(\omega_k t(\tau)-\bm{k}.\bm{x}(\tau))}$ in equation \eref{basic} is nonzero at the negative frequency $-\Delta E$. That this can happen is due to a mathematical phenomenon that may be called \it concomitant frequencies: \rm The Fourier transform of a wave that monotonically changes its frequency within a certain (e.g., negative) frequency interval contains also frequencies outside that interval (including, e.g., positive frequencies). 

To see explicitly that concomitant frequencies are responsible for the Unruh effect, let us consider the fact that $\bar{\omega}(\tau) >0$ at all times $\tau$ for all trajectories. This is because   
\bea
\bar{\omega}(\tau) = \frac{\rmd}{\rmd\tau}(\omega_k t(\tau)-\bm{k}.\bm{x}(\tau))=\eta_{\mu\nu}k^\mu\frac{\rmd}{\rmd\tau}x^\nu(\tau),
\eea
which is always positive as is easily seen in the instantaneous rest frames of the detector. Intuitively, the fact that $\bar{\omega}(\tau) >0~\forall \tau$ should prevent excitation of the detector since equation \eref{basic} shows that the detector's excitation probability is proportional to the value of the Fourier transform of $\e^{\ii \int_0^\tau\bar{\omega}(\tau')\rmd\tau'}$ at the negative frequency value $-\Delta E$. 
The reason why, nevertheless, the detector can become excited, i.e., the reason for why the probability amplitude $\psi_k(\Delta E)$ can be nonzero also for $\Delta E>0$ is because the Fourier transform, $\psi_k(\Delta E)$, generally shows the presence of both positive and negative frequencies in $\e^{\ii(\omega_k t(\tau)-\bm{k}.\bm{x}(\tau))}$ in spite of the fact that $\bar{\omega}(\tau)>0~\forall \tau$. 

\begin{figure}[htp!] 
\centering
	\includegraphics[width=0.5\textwidth]{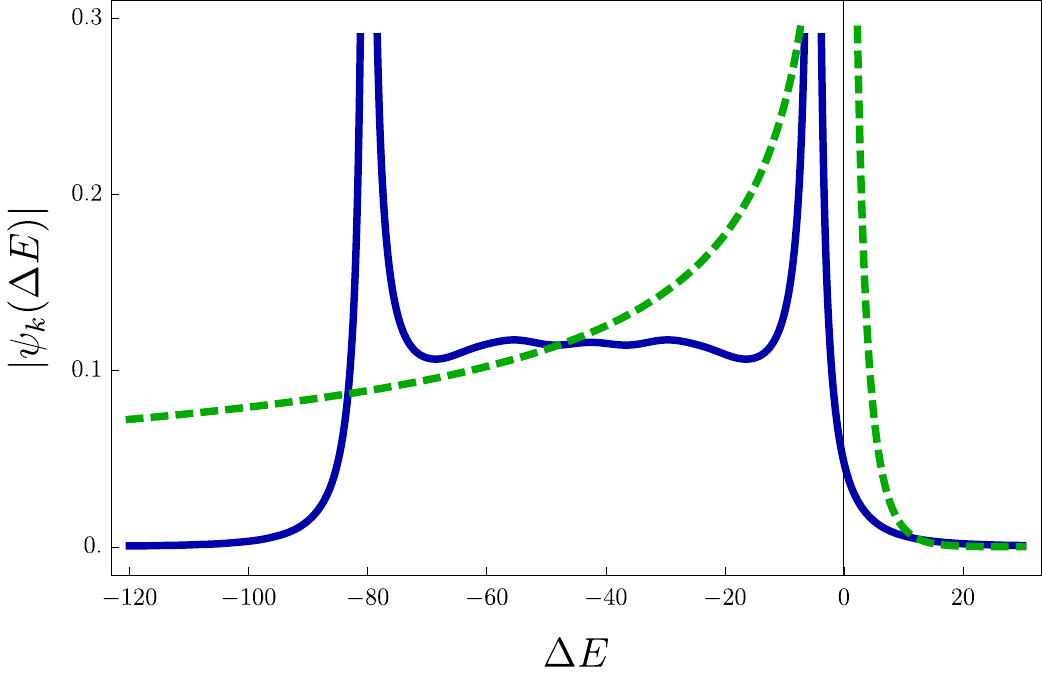}
  \caption{\label{fourier} The Fourier transform, $\vert\psi_k(\Delta E)\vert$, for a trajectory of uniform acceleration (dashed) and a trajectory that is inertial except for a period of temporary linear acceleration (solid) with the frequencies running through the interval $(5,80)$. The finite amplitudes on the positive axis show the concomitant frequencies that cause the Unruh effect.}
	\end{figure}

Let us consider, for example, the well-known case of a detector with constant acceleration, $a>0$, and trajectory $(t(\tau), {\bf x}(\tau))=(\sinh(a \tau)/a, \cosh(a \tau)/a,0,0)$ with $\bm{k}=(1,1,0,0)$ and $\chi(\tau)\equiv 1$. In this case, $\bar{\omega}(\tau) = k~\e^{-a\tau}$ which is positive for all $\tau$. Nevertheless, the Fourier integral equation \eref{basic}, which can be solved using the Gamma function, 
\begin{equation}
\psi_k(\Delta E) = \frac{\mu}{a}\sqrt{\frac{\Delta E}{\omega_k}} \rme^{-\frac{\pi \Delta E}{2a}}  \rme^{\ii\frac{\Delta E}{a}\ln(\frac{\omega_k}{a})}\Gamma(-\ii\Delta E/a)\label{gamma},
\end{equation}
shows that $\psi_k(\Delta E)$ contains both positive and negative frequencies, as shown by the dashed line in figure \ref{fourier}.

\bf Remark: \rm
While we here treat the Unruh effect using Unruh DeWitt detectors rather than using the alternative method of Bogolubov transformations, we remark that also the calculation of $\beta$ coefficients of Bogolubov transformations can be viewed as the calculation of generalized concomitant frequencies. For example, in the case of uniform acceleration, the same integral as that leading to equation \eref{gamma} arises in the calculation \cite{Mukhanov} of the Bogolubov $\beta$ coefficients. We note that, in general, the calculation of clicking probabilities of UDW detectors can differ from the calculation of the excitation probabilities of field modes, as is the case, for example, for circular motion, see, e.g., \cite{padmanabhan2002}. This is because the two methods are based on different definitions of particles, namely by identifying particles either as (local) UDW detector excitations or as (nonlocal) field mode excitations.

Another example is an always-on detector which, for a finite amount of time, is accelerated so that the frequency increases linearly, preceded and followed by inertial motion. In this case, therefore, we need to calculate the Fourier transform of the function $e^{\ii \int_0^\tau \bar{\omega}(\tau')\rmd\tau'}$ with $\bar{\omega}(\tau')$ being constant, then linearly increasing for a finite duration, then being constant again. This Fourier transform can be carried out straightforwardly because the nontrivial period of linear acceleration is a complex Gaussian integral on a finite interval which can be evaluated in terms of error functions. As shown with the solid curve in figure \ref{fourier}, also the Fourier transform $\vert\psi_k(\Delta E)\vert$ for this trajectory exhibits finite amplitudes for positive $\Delta E$ which cause the Unruh effect, i.e., it too exhibits the presence of concomitant frequencies: As one might expect, the spectrum $\vert\psi_k(\Delta E)\vert$ for the trajectory of temporary acceleration contains frequencies within the interval from the initial to the final frequency (on the negative half axis) including a peak at either end of the interval because the initial and final velocities are maintained for an infinite amount of time. In addition, the spectrum exhibits the presence of concomitant frequencies far into the negative and positive $\Delta E$ axes, similarly to how also the spectrum for the trajectory of uniform acceleration exhibits concomitant frequencies on the positive $\Delta E$ axis. 

\section{Trajectories that extremize the probability for Unruh processes}

Our aim now is to study the origin of the phenomenon of concomitant frequencies in equation \eref{basic} in order to determine how strong and how weak this phenomenon can be made, i.e., to what extent the probability amplitudes, $\psi_k(\Delta E)$, for Unruh processes can be modulated by choosing trajectories with suitable non-uniform accelerations. 

To this end, it will be convenient to define a function $\omega(\tau)$ such that the entire integrand, $u(\tau)$, in equation \eref{basic} takes the form 

\begin{equation} u(\tau) = \rme^{\ii \Delta E \tau} \rme^{\ii(\omega_k t(\tau)-\bm{k}.\bm{x}(\tau))}= \rme^{\ii \int_0^\tau \omega(\tau')~\rmd\tau'}.\end{equation} 
 We recall that, in an instantaneous rest frame, $x^\mu(\tau) =(\tau,0,0,0)$, so that $\rmd x^\mu(\tau)/\rmd \tau =\dot{x}^\mu=(1,0,0,0)$ implying $\dot{x}_\mu\dot{x}^\mu=1$ so that $\rmd(\dot{x}_\mu\dot{x}^\mu)/d\tau=0 = 2\ddot{x}_\mu\dot{x}^\mu$. Thus, $\dot{x}^\mu=(1,0,0,0)$ yields $\ddot{x}_0=\rmd^2t/\rmd\tau^2=0$ in an instantaneous rest frame. This implies that:
\bea 
\rmd\omega(\tau)/\rmd\tau= -{\bm k}.{\bm a}(\tau)=-\cos(\alpha)\vert{\bm k}\vert\vert{\bm a}\vert,
\eea
where ${\bm a}(\tau)$ is the acceleration 3-vector in the instantaneous rest frame and $\alpha$ is its angle with ${\bm k}$. Since $\vert {\bm a}\vert$ is proportional to the power that the observer spends on acceleration, the observer achieves the most change in $\omega(\tau)$ and therefore the strongest effect on $\psi_k(\Delta E)$, for a given energy budget, when accelerating parallel or antiparallel to the mode ${\bm k}$ considered and we will, therefore, focus on this case. 
However, before analyzing the production of concomitant frequencies in equation \eref{basic} further, we now separate off another effect which can also lead to detector excitations in the vacuum, namely the effect of a switching of the detector through a nontrivial function $\chi(\tau)$. To this end, we consider a detector that is inertial, which means that $\omega(\tau)$ is constant, $\omega(\tau)\equiv \omega$, with $\omega >0$. 
If, in this case, the detector is switched on for an integer number of cycles of the integrand, its integral $\psi_k(\Delta E)$ vanishes, $\int_0^{2\pi N/\omega} \rme^{\ii \omega \tau}\rmd\tau=0$. 
However, if the detector is kept on for a non-integer number of cycles (or if $\chi(\tau)$ is a generic smooth switching function), then $\psi_k(\Delta E)$ is finite. The effect is maximal for a half integer number of cycles, in which case $\vert\psi_k(\Delta E)\vert=2/\omega$.
Physically, the finite probability for an inertial detector to click due to this truncation effect expresses the time-energy uncertainty principle. 
Since our interest here is not in the time-energy uncertainty principle but in the phenomenon of concomitant frequencies we will, therefore, always consider an integer number of cycles.

The origin of concomitant frequencies is thereby traced to the fact that the integral over $N$ complete cycles does not need to be zero if the frequency $\omega(\tau)$ changes during the $N$ cycles. In the following, we will, therefore, consider the contribution, $\psi_k^{(N)}(\Delta E)$, of an integer number, $N$, of complete cycles of the integrand in equation \eref{basic}, to the probability amplitude, $\psi_k(\Delta E)$, of an always-on detector whose trajectory remains either parallel or antiparallel to ${\bm k}$ to maximize the effect. In this case, $\rmd\omega(t)/\rmd\tau= -{\bm k}.{\bm a}(\tau)$ has a definite sign and 
$\omega(\tau')$ is monotonically either decreasing or increasing. We denote the time interval of the $N$ cycles by $[0,\tau_f]$, i.e., we have $u(0)=u(\tau_f)=1$.  

We arrive at the basic extremization problem to find those monotonic functions, $\omega(\tau)$, obeying say $0<\omega(0)<\omega(\tau_f)$ which extremize

\begin{equation}
\vert \psi_k^{(N)}(\Delta E)\vert = \left\vert\int_0^{\tau_f} \rme^{\ii \int_0^\tau \omega(\tau') \rmd\tau'}\rmd\tau\right\vert, \label{two}
\end{equation}
while completing $N$ cycles. From $\omega(\tau)$ one then obtains via $\rmd\omega(t)/\rmd\tau= -{\bm k}.{\bm a}(\tau)$ the proper accelerations ${\bm a}(\tau)$ and therefore the desired piece of trajectory which contributes extremally to $\psi_k(\Delta E)$ for fixed $\bm{k}, \Delta E$, $N$ and for fixed initial and final velocities.

We can solve the corresponding constrained variational problem by using symmetry considerations. To this end, we change the integration variable in equation \eref{two} by defining $v(\tau):=\int_0^\tau \omega(\tau')\rmd\tau'$, so that $\rmd\tau/\rmd v = 1/\omega(\tau(v))$:  
\begin{equation}
\psi_k^{(N)}(\Delta E) = \int_0^{2\pi N} \rme^{\ii v} \frac{1}{\omega(\tau(v))}~\rmd v.
\end{equation}
The extremization problem is now to find those monotonic functions $\omega(\tau(v))$ which extremize $\vert\psi_k^{(N)}(\Delta E)\vert$. The integral is equivalent to calculating the center of mass of a wire of length $2\pi N$ coiled up $N$ times on the unit circle in the complex plane with the wire's mass density at length $v$ being $1/\omega(\tau(v))$. The problem of maximizing $\vert\psi_k^{(N)}(\Delta E)\vert$ is now to monotonically vary the mass density of the wire  between its prescribed initial and final values $1/\omega(0)$ and $1/\omega(\tau(2\pi N))$ such that the center of mass of the coiled-up wire is as much as possible off center. 

For $N=1$, the answer is clearly to put as much mass as possible on one half circle and as little as possible on the other half. This means that $\vert\psi_k^{(1)}(\Delta E)\vert$ is maximal if the initial $\omega$ is maintained, $\omega(\tau(\nu)) \equiv \omega(0)$, in the first half of the cycle and then the frequency is abruptly changed to the final frequency which is then maintained, $\omega(\tau(\nu)) \equiv \omega(2\pi)$, for the second half of the cycle. All acceleration happens abruptly mid cycle. The maximum is, therefore:

\begin{equation}
\vert\psi_k^{(1)}(\Delta E)\vert = \left\vert\int_0^{2\pi} \frac{\e^{\ii v}}{\omega(\tau(v))}~\rmd v\right\vert = 2\left\vert\frac{1}{\omega(0)}-\frac{1}{\omega(\tau_f)}\right\vert. \label{max}
\end{equation}
For $N>1$, by the same argument, it is optimal to pursue this acceleration regime in one of the cycles and to have all prior and subsequent cycles at constant velocity.  
Equation \eref{max} also shows that the effect of concomitant frequencies can grow as large as  $2\max(1/\omega(0),1/\omega(\tau(2\pi)))$ which is the maximal size of the truncation effect due to the time-energy uncertainty principle. 

Conversely, there are trajectories over $N$ cycles that contribute minimally to $\vert\psi_k(\Delta E)\vert$. These are the trajectories where all accelerations are abrupt and occur only at the beginnings (or ends) of cycles. In this way, all cycles are individually monochromatic and do not contribute. The minimum is, therefore, $\vert\psi_k^{(N)}(\Delta E)\vert =0$.  

We conclude that suitable non-uniform acceleration over $N$ cycles is able to modulate the amplitude, $\vert\psi^{(N)}_k(\Delta E)\vert$, for individual Unruh processes within the range $(0,2\left\vert 1/\omega(0)-1/\omega(\tau_f)\right\vert)$. 

We remark that also for periodic trajectories, by the same reasoning, sudden accelerations, e.g., up and down, are optimal. In this case, the accelerated UDW detector acts like an electron moving up and down an antenna with constant velocities ${\bm v_u}$ and ${\bm v_d}=-{\bm v_u}$ with corresponding frequencies $\omega_u$ and $\omega_d$, respectively. From $\omega(\tau) = \Delta E +\omega_k\rmd t/\rmd \tau - {\bm k}. {\bm v(\tau)}$ with ${\bm v(\tau)} = d{\bm x}/d\tau$, we have  $\omega_u = \Delta E +\omega_k \rmd t/\rmd \tau - {\bm k}. {\bm v_u}$ and 
$\omega_d = \Delta E +\omega_k \rmd t/\rmd \tau + {\bm k}. {\bm v_d}$. 
Let us now determine the optimal frequency $f_{osc}$ with which the UDW detector is to move up and down (with sudden accelerations at both ends) to maximize $\vert\psi^{(1)}_k(\Delta E)\vert$. From the picture of the wire above, we know that the frequencies $\omega_u$ and $\omega_d$ are to last for a half period each. We conclude that $2\pi/f_{osc} = \pi/\omega_u+\pi/\omega_d$, and that, therefore, the optimal frequency, $f_{osc}$, at which the UDW detector should oscillate up and down to maximize $\vert\psi^{(1)}_k(\Delta E)\vert$ is the harmonic mean: $f_{osc}=2/(1/\omega_u+1/\omega_d)$. Interestingly, this analog of an antenna resonance frequency, $f_{osc}$, is not simply determined by the frequency, $\omega_k$, of the field quantum emitted in the Unruh process, as one might have expected from the antenna analogy (in the nonrelativistic limit where $t=\tau$). Instead, as 
$\omega(\tau) = \Delta E +\omega_k\rmd t/\rmd\tau - {\bm k}. {\bm v(\tau)}$ shows, $f_{osc}$ is larger because it also contains a positive contribution from $\Delta E$. Intuitively, this is because for an Unruh process to occur, sufficient energy is required not only to emit a field quantum, as from an antenna, but also to excite the UDW detector from its ground state to its excited state. Further, from the picture of the wire above, we notice that the more different the two frequencies $\omega_u$ and $\omega_d$ are, i.e., the larger the amplitude of the UdW detector's motion, the larger is 
$\vert \psi^{(1)}_k(\Delta E)\vert$. Interestingly, the cosine term in ${\bm k.v}(\tau)$, therefore, also implies that the emission of the scalar field quanta by an UDW detector is mostly in the direction of motion, unlike in the case of the emission of the always transversally-polarized photons from antennas.

\section{Optimal regularized trajectories}
The above results confirm the expectation that suitable nonzero higher-than-second derivatives in trajectories can significantly modulate the Unruh amplitudes $\vert\psi_k^{(N)}(\Delta E)\vert$. However, these trajectories are unrealistic in the sense that they are distributional by requiring sudden arbitrarily large higher derivatives. The question arises to what extent $\vert\psi^{(N)}_k(\Delta E)\vert$ can be modulated by realistic trajectories that are regular in the sense that they possess only a few significantly nonzero higher derivatives. Further, the question is to what extent such trajectories can enhance the Unruh effect when compared to trajectories of uniform acceleration which possess the same initial and final velocities, i.e., that possess the same energy budget for acceleration.

In order to calculate those trajectories that optimize $\vert\psi_k^{(N)}(\Delta E)\vert$ when allowing only finitely many derivatives to be nonzero, we solved the problem of constrained optimization of $\vert\psi_k^{(N)}(\Delta E)\vert$ by finding a suitable trajectory numerically. We optimized among trajectories whose proper acceleration functions, $\omega(\tau)=-{\bm k}.{\bm a}(\tau)$, are polynomials of a pre-determined maximal degree, and which are monotonically increasing between fixed initial and final values, within a fixed integer number, $N$, of cycles. 
\begin{figure}[h] 
\centering
	\includegraphics[width=0.5\textwidth]{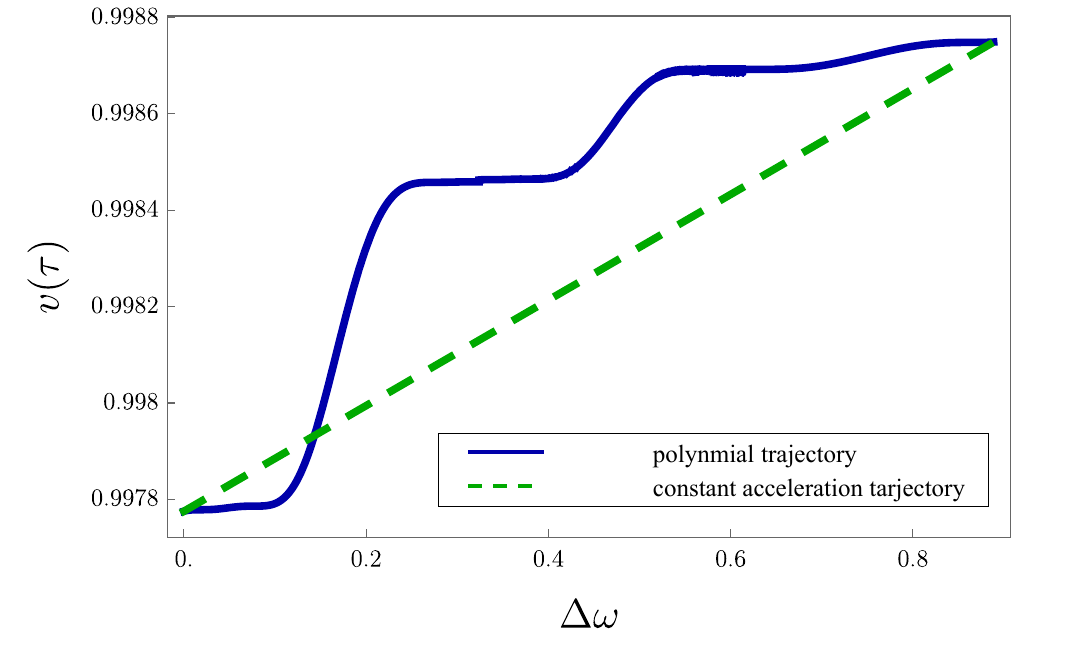}
  \caption{\label{traj3cyc}The velocity, $v(\tau)$, for a straight line trajectory of uniform acceleration (dashed), and the velocity $v(\tau)$ of a polynomial straight line  trajectory that is orthogonal to $\bm{k}$,  of degree 13 and optimizes $\vert\psi_k^{(3)}(\Delta E)\vert$  over three cycles (solid).}
	\end{figure}
Figure \ref{traj3cyc} compares a trajectory of constant acceleration with the trajectory that maximizes $\vert\psi_k^{(N)}(\Delta E)\vert$ among all trajectories of polynomial degree $n\le 13$ for $N=3$ cycles and the same overall change in velocity. We notice that the latter trajectory involves alternating periods of diminished and enhanced acceleration. We can now address to what extent a trajectory with non-uniform acceleration can increase the probability for an elementary Unruh process to occur. 
	
\begin{figure}[htp] 
	\includegraphics[width=0.45\textwidth]{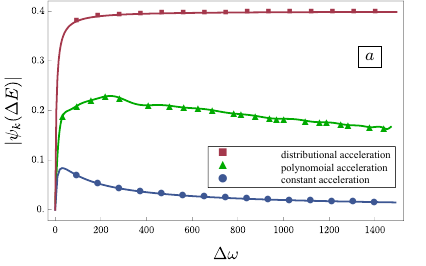}
	\includegraphics[width=0.45\textwidth]{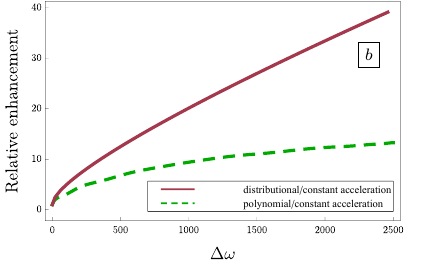}
  \caption{\label{ratio}(a) Contribution to $\vert\psi_k(\Delta E)\vert$ by one cycle, as a function of $\Delta \omega$ for a trajectory of constant acceleration (circles), for a polynomial trajectory (triangles), and for a trajectory with a distributional acceleration (squares). (b) Relative enhancement of elementary Unruh processes: the two curves show the ratio of the two upper curves by the lowest curve of figure 3(a). Here, the initial frequency is $\omega(0)=5$ and the energy gap of the detector is $\Delta E=3$}.
	\end{figure}

To this end we compare, see figure \ref{ratio}, the moduli of the probability amplitudes $\vert\psi_k^{(N)}(\Delta E)\vert$ for elementary Unruh processes for three types of trajectories with the same initial and final velocities: an optimal trajectory (which is of distributional acceleration, i.e., all fuel available for acceleration is burned at once), a regularized optimal trajectory of non-uniform acceleration that is polynomial of a given degree, and for a trajectory of uniform acceleration. We see that the larger the difference between the initial and final velocities, the more $\vert\psi_k^{(N)}(\Delta E)\vert$ can be enhanced by a trajectory of non-uniform acceleration.  
	
\section{The role of non-adiabaticity in trajectories}
We investigated the extent to which the probability for an individual Unruh process can be enhanced by choosing a trajectory of suitable non-uniform acceleration while holding fixed the initial and final velocities and the number of cycles. We found that the probability for an Unruh process can be enhanced strongly over its probability for a trajectory of uniform acceleration, namely with trajectories whose acceleration changes suddenly. $\vert\psi^{(N)}_k(\Delta E)\vert$ can reach as high as $2\left\vert 1/\omega(0)-1/\omega(\tau_f)\right\vert$, whose magnitude is comparable to $\max(2/\omega(0),2/\omega(\tau_f))$, which is the maximal size of the effect due to a sudden switching of the detector. To see that this is a strong effect, let us recall that  the effect of sudden switching of an UDW detector is known to be large in the sense that in 3+1 dimensional Minkowski space the cumulative first-order excitation probability obtained by integrating $\vert\psi_k(\Delta E)\vert^2$ over all modes $\bm{k}$, diverges \cite{ottewill2002,Louko2008,Satz2006}.    

We also found that the change of acceleration of a trajectory does not need to be entirely sudden for the strong modulation to set in. By analyzing the origin of concomitant frequencies, we found that the condition for the modulation of $\vert\psi_k(\Delta E)\vert$ to be strong is that there is a significant change in frequency within one cycle, i.e, that the trajectory is non-adiabatic in this sense. The Unruh temperature for a trajectory of uniform acceleration, $a=|\bm{a}|$, is $T=a/2\pi$, with the dominant radiation at a wavelength of $\lambda \approx 1/a$, in natural units. In practice, therefore, it should be possible to strongly modulate the dominant Unruh processes, and perhaps bring them closer to observability, if one can significantly change the momentary acceleration $\bm{a}(t)$ at or below the time scale $\lambda \approx 1/a(t)$. We conjecture that, similarly, also the dynamical Casimir effect, see, e.g., \cite{wilson2011observation}, can be enhanced through suitably non-uniformly accelerated trajectories. 


\section{Reconstruction of trajectories from Unruh processes} 
Having established the extent to which nonuniform acceleration modulates Unruh processes, we now show that the Unruh spectrum $\psi_k(\Delta E)$ (for all $\Delta E\in \mathbb{R}$ and ${\bm k}\in \mathbb{R}^3$) contains enough information to reconstruct the trajectory. To see this, we Fourier transform, take a log and differentiate equation \eref{basic}:
\be
\frac{\rmd}{\rmd k_i}\ln \left[\int_{\mathbb{R}} \psi_k(\Delta E) \rme^{-\ii\tilde{\tau}\Delta E}\rmd \Delta E\right]=\frac{\ii~k_i~ t(\tilde{\tau})}{\sqrt{{\bm k}^2 +m^2}} -\ii x_i(\tilde{\tau}).
\ee
The ambiguity in the logarithm's branch cuts is fixed by the initial condition and by the continuity of the right hand side. 
Since $t(\tilde{\tau})$ is determined by the $x_i(\tilde{\tau})$, in order to obtain the trajectory $x_i(\tilde{\tau})$, it suffices to evaluate this expression at one value for ${\bm k}$, for example, for ${\bm k}=0$:

\be
x_i(\tau) = \ii \frac{\rmd}{\rmd k_i}\ln \left[\int_{\mathbb{R}} \psi_k(\Delta E) \rme^{-\ii{\tau}\Delta E}\rmd \Delta E\right]_{{\bm k}=0} .
\ee
As is well-known, an observer who knows that she is  uniformly accelerated can measure the magnitude of her acceleration by measuring the Unruh temperature. 
Here, we found that if an observer travels on an arbitrary unknown trajectory and possesses an ensemble of UDW detectors with both positive and negative gaps (i.e., that start in the ground and excited states, respectively), and if she keeps track of the emitted Unruh quanta at just one point in the spectrum (e.g., ${\bm k} = 0$) then from the Unruh spectrum $\psi_k(\Delta E)$ she can, in principle, fully reconstruct her trajectory. 

\ack
AK acknowledges support through the Discovery Grant Program of the Canadian National Science and Engineering Research Council (NSERC). AA and AK acknowledge helpful feedback from E. Mart\'{\i}n-Mart\'{\i}nez, R. B. Mann and J. Louko.

\section*{References}
\bibliographystyle{unsrt}
\bibliography{cavity_refs}

\end{document}